\documentclass[11pt]{article}
\usepackage{amssymb}
\usepackage{colortbl}
\usepackage{amsfonts,amsmath, longtable}

        \def\theequation{\thesection.\arabic{equation}}

\newcommand{\tr}{{\rm tr}}
\newcommand{\ti}[1]{\tilde{#1}}

\newcommand{\mH}{{\mathcal H}}

\newcommand{\al}{\alpha}

\newcommand{\Mat}{ {\rm Mat}(N,\mathbb C) }

\newcommand{\mC}{\mathbb C}

\def\beq{\begin{equation}}
\def\eq{\end{equation}}
\def\p{\partial}

\begin{document}

\setcounter{page}{1}

\begin{center}

\

\vspace{-0mm}

{\Large{\bf Gauge equivalence of 1+1 Calogero-Moser-Sutherland field theory }}

\vspace{3mm}

{\Large{\bf and higher rank trigonometric Landau-Lifshitz model }}

%


 \vspace{15mm}

 {\Large {K. Atalikov}}$\,^{\bullet}$
\qquad\quad\quad
 {\Large {A. Zotov}}$\,^{\diamond\,\bullet}$

  \vspace{5mm}

$\diamond$ -- {\em Steklov Mathematical Institute of Russian
Academy of Sciences,\\ Gubkina str. 8, 119991, Moscow, Russia}


$\bullet$ -- {\em NRC ''Kurchatov Institute'',\\
Kurchatova sq. 1, 123182, Moscow, Russia}


   \vspace{3mm}

 {\small\rm {e-mails: kantemir.atalikov@yandex.ru, zotov@mi-ras.ru}}

\end{center}

\vspace{0mm}

\begin{abstract}
We consider the classical integrable 1+1 trigonometric ${\rm gl}_N$ Landau-Lifshitz models
 constructed by means of quantum $R$-matrices satisfying
also the associative Yang-Baxter equation. It is shown that
1+1 field analogue of the trigonometric Calogero-Moser-Sutherland model
is gauge equivalent to the Landau-Lifshitz model  which arises
from the Antonov-Hasegawa-Zabrodin trigonometric non-standard $R$-matrix. The latter
generalizes the Cherednik's 7-vertex $R$-matrix in ${\rm GL}_2$ case to the case of ${\rm GL}_N$.
Explicit change of variables between the 1+1 models is obtained.
\end{abstract}

%

{\small{
\tableofcontents
}}


\newpage

\section{Introduction}
\setcounter{equation}{0}

In this paper we consider two types of models arising as 1+1 field generalizations of classical integrable
finite-dimensional systems. At the level of finite-dimensional mechanics the
 first one is the Calogero-Moser-Sutherland many-body system \cite{Calogero}.
In the trigonometric case\footnote{We do not distinguish trigonometric and hyperbolic models
since all variables are complex-valued.} it is defined by the Hamiltonian
 \beq\label{a01}
 \begin{array}{c}
  \displaystyle{
 H^{\hbox{\tiny{CMS}}}=\frac{1}{2}\sum\limits_{i=1}^N \frac{p_i^2}{2}-\sum\limits_{i<j}\frac{c^2}{4\sinh^2(\frac{q_i-q_j}{2})}\,,
  }
 \end{array}
\eq
 where $p_i$ and $q_i$, $i=1,...,N$ are canonically conjugated momenta and positions of particles
 \beq\label{a011}
 \begin{array}{c}
  \displaystyle{
\{q_i,p_j\}=\delta_{ij}\,,\qquad \{p_i,p_j\}=\{q_i,q_j\}=0
  }
 \end{array}
\eq
 and $c\in\mC$ is a coupling constant.
 The second type model at the level of classical mechanics is an integrable trigonometric top \cite{KrZ} of
 the Euler-Arnold type \cite{Arnold}. The latter means that the Hamiltonian is given as
 \beq\label{a02}
 \begin{array}{c}
  \displaystyle{
 H^{\hbox{\tiny{top}}}=\frac{1}{2}\,\tr(SJ(S))\,,\qquad S\in\Mat\,,
  }
 \end{array}
\eq
 where the matrix elements of $S$ are dynamical variables and $J(S)$ is some special linear functional.
 Together with the Poisson-Lie brackets on ${\rm gl}^*_N$ Lie coalgebra
 \beq\label{a021}
 \begin{array}{c}
  \displaystyle{
 \left\{S_{i j}, S_{k l}\right\}=\frac{1}{N}\,\Big(S_{i l} \delta_{k j}-S_{k j} \delta_{i l}\Big)
  }
 \end{array}
\eq
 the Hamiltonian (\ref{a02}) provides equations of motion
 in the Euler-Arnold form
 \beq\label{a03}
 \begin{array}{c}
  \displaystyle{
 \{S,H^{\hbox{\tiny{top}}}\}\equiv{\dot S}=[S,J(S)]\,.
  }
 \end{array}
\eq
 Both models (\ref{a01}) and (\ref{a02}) are described by the Lax equations
 \beq\label{a04}
 \begin{array}{c}
  \displaystyle{
 {\dot L}(z)=[M(z),L(z)]\,,\quad L(z),M(z)\in\Mat\,,
  }
 \end{array}
\eq
 where $z\in\mC$ is a spectral parameter. The dimension of the phase space of the integrable top
 depends on the choice of the coadjoint orbit, i.e. on some fixation of eigenvalues of $S$ which are the Casimir functions of (\ref{a021}). In the case of the coadjoint orbit of minimal dimension (when $N-1$ eigenvalues of $S$ coincide)
 the phase space has dimension $2N-2$, which is equal to the one for the Calogero-Moser-Sutherland model in the center
 of mass frame. In this particular case (and for some special $J(S)$) two models can be shown to be gauge equivalent, that is there exist a
 gauge transformation matrix $g(z)=g(z,q_1,...,q_N)\in\Mat$, which maps one Lax matrix to another one:
 \beq\label{a05}
 \begin{array}{c}
  \displaystyle{
 L^{\hbox{\tiny{top}}}(z)=g(z)L^{\hbox{\tiny{CMS}}}(z)g^{-1}(z)\,.
  }
 \end{array}
\eq
 This allows to compute explicit change of variables $S=S(p,q,\nu)$, which provides the Poisson canonical map
 between phase spaces of both models endowed with the Poisson brackets (\ref{a011}) and (\ref{a021}) respectively.
 The gauge transformation (\ref{a05}) can be considered as the classical analogue of the IRF-Vertex correspondence
 between dynamical and non-dynamical quantum $R$-matrices:
 \cite{Baxter2}:
  \beq\label{a06}
 \begin{array}{c}
  \displaystyle{
R^{\hbox{\tiny{Vertex}}}_{12}(\hbar,z_1-z_2)
  =
g_2(z_2,q)\,
g_1(z_1,q-\hbar^{(2)})\,R^{\hbox{\tiny{IRF}}}_{12}(\hbar,z_1-z_2|\,q)
g^{-1}_2(z_2,q-\hbar^{(1)})g_1^{-1}(z_1,q)\,,
}
\\ \ \\
  \displaystyle{
g_1(z_1,q-\hbar^{(2)})=P_2^{-\hbar}g_1(z_1,q)P_2^\hbar\,,\quad
P_2^\hbar=\sum\limits_{k=1}^N  1_N\otimes E_{kk}
\exp(\hbar\p_{q_k})\,,
 }
 \end{array}
 \eq
 where $g(z,q)=g(z,q_1,...,q_N)\in\Mat$ is some special matrix (the intertwining matrix) providing the IRF-Vertex transformation.
 The matrix $g(z)$ from (\ref{a06}) is, in fact, exactly the one, which is used in
 (\ref{a05}).
 The non-dynamical (vertex type) $R$-matrix, by definition, satisfies the quantum Yang-Baxter equation:
 \beq\label{a07}
 \begin{array}{c}
  \displaystyle{
 R_{12}^\hbar R_{13}^\hbar R_{23}^\hbar
=R_{23}^\hbar R_{13}^\hbar R_{12}^\hbar\,,
   \qquad R^\hbar_{ab} = R^{\hbox{\tiny{Vertex}}}_{ab}(\hbar,z_a-z_b)\,.
  }
 \end{array}
\eq
 The classical IRF-Vertex relation between many-body systems and top-like models
 was proposed in \cite{LOZ} (see also \cite{AASZ,LOZ14}) in the framework of Hitchin approach to integrable systems.
 It was called the symplectic Hecke correspondence since it changes a certain characteristic class
 of underlying bundles. For the trigonometric $R$-matrices in ${\rm GL}_N$ case the relation
 (\ref{a06}) was described in \cite{AHZ},
 and the vertex-type $R$-matrix was evaluated
 in the special trigonometric limit starting from the elliptic $R$-matrix. The resultant
 $R$-matrix was called the non-standard trigonometric $R$-matrix. It generalizes the Cherednik's
 7-vertex $R$-matrix \cite{Chered} in ${\rm GL}_2$ case. At the classical level the relation
  (\ref{a06}) was described in \cite{KrZ}.

 {\bf The purpose of the paper} is to describe the IRF-Vertex type relation between
 the classical 1+1 field generalizations of the trigonometric models (\ref{a01}) and (\ref{a02}).
 In 1+1 case the Lax equation (\ref{a04}) turns into the Zakharov-Shabat equation
 written for the fields depending on the time variable $t$ and the space variable $x$:
 \beq\label{a08}
 \begin{array}{c}
  \displaystyle{
 \partial_{t}{U}(z)-k\partial_{x}{V}(z)+[{U}(z), {V}(z)]=0\,,\qquad {U}(z), {V}(z)\in\Mat\,,
  }
 \end{array}
\eq
 where $k\in\mC$ is a constant parameter. The limit to the finite-dimensional mechanics (when all fields are independent of $x$) corresponds to $k\rightarrow 0$. Then the zero curvature equation (\ref{a08})
 becomes the Lax equation (\ref{a04}). The field analogue for integrable finite-dimensional many-body
 systems was first proposed by A. Mikhailov for the Toda model \cite{Mikh}. The particles momenta and positions
 of particles become fields with the Poisson brackets
 \beq\label{a09}
 \begin{array}{c}
  \displaystyle{
\{q_i(x),p_j(y)\}=\delta_{ij}\delta(x-y)\,,\qquad \{p_i(x),p_j(y)\}=\{q_i(x),q_j(y)\}=0\,.
  }
 \end{array}
\eq
 Hereinafter we assume that all fields are periodic functions on a circle, i.e. $q_i(x+2\pi)=q_i(x)$
 and similarly for all other fields.
 The field generalization for
 the Calogero-Moser-Sutherland model was introduced in \cite{Krich2} and \cite{LOZ}. In paper \cite{Krich22}
 the final explicit description was suggested for ${\rm sl}_N$ (i.e. $N$-body) case.
 Let us also mention the field analogue of the Ruijsenaars-Schneider model \cite{ZZ}, which
 generalizes results of  \cite{Krich22} to semi-discrete Zakharov-Shabat equations.

 The filed generalization of the top-like models is given by the Landau-Lifshitz type models
 of 1-dimensional magnets \cite{LL}. In ${\rm sl}_2$ case equation of motion takes the form
  \beq\label{a10}
  \begin{array}{l}
  \displaystyle{
 \p_t {S}= [{S}, J({S})]+ [{S}, \p_x^2{S}]\,,\quad S=S(t,x)\in{\rm Mat}(2,\mC)\,.
 }
 \end{array}
 \eq
 Integrability of the latter model was proved in \cite{Skl}. For our purpose we need
 some higher rank ${\rm gl}_N$ extension of the Landau-Lifshitz equation.
  One possible higher rank generalization of (\ref{a10}) was proposed in \cite{GS}. In this paper
  we deal with another construction for ${\rm gl}_N$ case suggested in \cite{AtZ2}.
It is based on the associative Yang-Baxter equation \cite{FK}:
 \beq\label{a11}
 \begin{array}{c}
  \displaystyle{
 R^{\hbar}_{12} R^{\eta}_{23} = R^{\eta}_{13} R^{\hbar-\eta}_{12} + R^{\eta-\hbar}_{23} R^{\hbar}_{13}\,,
   \qquad R^x_{ab} = R^x_{ab}(z_a-z_b)
  }
 \end{array}
\eq
written for the vertex type $R$-matrices. Namely, it was shown in \cite{AtZ2} that the
coefficients of expansion of $R$-matrix satisfying (\ref{a11}) can be used for construction
of the $U$-$V$ pair with spectral parameter providing some higher rank generalizations
of (\ref{a10}) through the Zakharov-Shabat equation (\ref{a08}). Classification of
trigonometric ${\rm GL}_N$ solutions of (\ref{a11}) was suggested in \cite{Pol2,T}.
It includes (the properly normalized) non-standard $R$-matrix from \cite{AHZ}.
We use this $R$-matrix and apply the construction from \cite{AtZ2} to define
the higher rank Landau-Lifshitz model. Then we show that there exists a gauge transformation $G\in\Mat$, which
relates it with the 1+1 Calogero-Moser-Sutherland model via the gauge transformation:
 \beq\label{a12}
 \begin{array}{l}
  \displaystyle{
 U^{\hbox{\tiny{LL}}}(z)=G(z)U^{\hbox{\tiny{2dCMS}}}(z)G^{-1}(z)+k\p_x G(z)G^{-1}(z)\,,
  }
  \\ \ \\
    \displaystyle{
 V^{\hbox{\tiny{LL}}}(z)=G(z)V^{\hbox{\tiny{2dCMS}}}(z)G^{-1}(z)+\p_t G(z)G^{-1}(z)\,.
  }
 \end{array}
\eq
 The existence of the relation (\ref{a12}) was argued in \cite{LOZ}, and the explicit changes of variables were
 found in \cite{AtZ1} for ${\rm sl}_2$ models (in the rational, trigonometric and elliptic cases) and in \cite{AtZ3}
 for ${\rm sl}_N$ rational models. In this resect the aim of this paper is to extend the results of
 \cite{AtZ3} to the trigonometric models.
 
 The phenomenon of gauge equivalence is known for other models. For example, the equivalence
  exists between 1+1  Heisenberg magnet
and the nonlinear Schrodinger equation \cite{ZaTa}. 
Another example is the gauge transformation with the classical
$r$-matrix structure in WZNW (and Toda) theories described in \cite{Feher}.

\section{Antonov-Hasegawa-Zabrodin non-standard $R$-matrix}
\setcounter{equation}{0}

Any non-dynamical $R$-matrix in the fundamental representation of ${\rm GL}_N$ is represented in the form
 \beq\label{a13}
 \begin{array}{c}
  \displaystyle{
 R_{12}^\hbar(z_1,z_2)=\sum\limits_{i,j,k,l=1}^N R_{ij,kl}(\hbar,z_1,z_2)E_{ij}\otimes E_{kl}\in\Mat^{\otimes 2}\,,
  }
 \end{array}
\eq
 where $z_1,z_2$ are spectral parameters, $\hbar$ is the Planck constant, the set $E_{ij}\in\Mat$, $i,j=1,...,N$ is the basis
 of matrix units in $\Mat$ and $R_{ij,kl}(\hbar,z_1,z_2)$ is a set of functions.
 By definition of a quantum $R$-matrix, it satisfies the quantum Yang-Baxter equation (\ref{a07}).
 In what follows we also assume
 \beq\label{a14}
 \begin{array}{c}
  \displaystyle{
 R_{12}^\hbar(z_1,z_2)=R_{12}^\hbar(z_1-z_2)\,.
  }
 \end{array}
\eq
 In \cite{AHZ} the following $R$-matrix (up to a choice of normalization coefficients) was computed by a special
 trigonometric limit:
 \beq\label{a15}
 \begin{array}{c}
  \displaystyle{
R_{i j, k l}^{\eta}(z)=\delta_{i j} \delta_{k l} \delta_{i k} \frac{N}{2}(\operatorname{coth}(N z / 2)+\operatorname{coth}(N \eta / 2))+
 }
 \\ \ \\
 \displaystyle{
+\delta_{i j} \delta_{k l} \varepsilon(i \neq k) \frac{N e^{(i-k) \eta-\operatorname{sgn}(i-k) N \eta / 2}}{2 \sinh (N \eta / 2)}+\delta_{i l} \delta_{k j} \varepsilon(i \neq k) \frac{N e^{(i-k) z-\operatorname{sgn}(i-k) N z / 2}}{2 \sinh (N z / 2)}+
 }
 \\ \ \\
 \displaystyle{
+N \delta_{i+k, j+l} e^{(i-j) z+(j-k) \eta}(\varepsilon(i<j<k)-\varepsilon(k<j<i))+
 }
 \\ \ \\
 \displaystyle{
+N e^{-N \Lambda} \delta_{i+k, j+l+N}\left(\delta_{i N} e^{-j z-l \eta}-\delta_{k N} e^{l z+j \eta}\right)\,,
  }
 \end{array}
\eq
 where $\Lambda\in\mC$ is a free constant.
It is called the non-standard trigonometric $R$-matrix. Here the following notation is used:
  \beq\label{a16}
  \begin{array}{c}
  \displaystyle{
 \varepsilon(\hbox{A})=\left\{\begin{array}{l} 1\,,\hbox{if A is true}\,,\\ 0\,,\hbox{if A is false}\,.\end{array}\right.
 }
 \end{array}
 \eq
 In the $N=2$ case it is the 7-th vertex $R$-matrix proposed by I. Cherednik \cite{Chered}:
  \beq\label{a161}
   \begin{array}{c}
  R^\hbar(z)=\left(\begin{array}{cccc} \coth(z)+\coth(\hbar) & 0 & 0 & 0\vphantom{\Big|}
  \\ 0 & \sinh^{-1}(\hbar) & \sinh^{-1}(z) & 0\vphantom{\Big|}
  \\ 0 & \sinh^{-1}(z) & \sinh^{-1}(\hbar) & 0\vphantom{\Big|}
  \\ -4\,e^{-2\Lambda}\sinh(z+\hbar) & 0 & 0 & \coth(z)+\coth(\hbar)          \vphantom{\Big|} \end{array} \right)\,.
  \end{array}
  \eq
 The $R$-matrix (\ref{a15}) satisfies the skew-symmetry
  \beq\label{a17}
  \begin{array}{c}
  \displaystyle{
 R^\hbar_{12}(z)=-R_{21}^{-\hbar}(-z)=-P_{12}R_{12}^{-\hbar}(-z)P_{12}\,,
 \qquad
 P_{12}=\sum\limits_{i,j=1}^N E_{ij}\otimes E_{ji}\,,
 }
 \end{array}
 \eq
 where $P_{12}$ is the matrix permutation operator. Also, it satisfies the
unitarity property
 \beq\label{a18}
   \begin{array}{c}
 \displaystyle{
R^\hbar_{12}(z) R^\hbar_{21}(-z) = f^\hbar(z)\,\,1_N\otimes 1_N
 }
  \end{array}
  \eq
  with the normalization factor
  \beq\label{a19}
   \begin{array}{c}
   \displaystyle{
  f^\hbar(z)=\frac{N^2}{4}\left(\frac{1}{\sinh^2(N\hbar/2)}-\frac{1}{\sinh^2(Nz/2)}\right)
 }
  \end{array}
  \eq
and the following set of local expansions:
  \beq\label{a20}
  \begin{array}{c}
      \displaystyle{
R^\hbar_{12}(z)=\frac{1}{\hbar}\,1_N\otimes 1_N+r_{12}(z)+\hbar\,
m_{12}(z)+O(\hbar^2)\,, }
  \end{array}
  \eq
  \beq\label{a21}
  \begin{array}{c}
      \displaystyle{
R^\hbar_{12}(z)=\frac{1}{z}\,P_{12}+R^{\hbar,(0)}_{12}+O(z)\,,
}
  \end{array}
  \eq
  \beq\label{a22}
  \begin{array}{c}
      \displaystyle{
 R^{\hbar,(0)}_{12}=\frac{1}{\hbar}\,1_N\otimes
 1_N+r^{(0)}_{12}+O(\hbar)\,,\qquad
 r_{12}(z)=\frac{1}{z}\,P_{12}+r^{(0)}_{12}+O(z)\,.
 }
  \end{array}
  \eq
The skew-symmetry (\ref{a17}) yields
  \beq\label{a23}
  \begin{array}{c}
  r_{12}(z)=-r_{21}(-z)\,,\qquad
  m_{12}(z)=m_{21}(-z)\,,\\ \ \\
      R^{\hbar,(0)}_{12}=-R^{-\hbar,(0)}_{21}\,,\qquad
  r_{12}^{(0)}=-r_{21}^{(0)}\,.
  \end{array}
  \eq
The coefficient $r_{12}(z)$ from (\ref{a20}) is the classical $r$-matrix:
  \beq\label{a24}
  \begin{array}{c}
      \displaystyle{
r_{i j, k l}(z)=\delta_{i j} \delta_{k l} \delta_{i k} \frac{N}{2} \operatorname{coth}(N z / 2)+
 }
 \\ \ \\
 \displaystyle{
+\delta_{i j} \delta_{k l} \varepsilon(i \neq k)\left((i-k)-\frac{N \operatorname{sgn}(i-k)}{2}\right)+\delta_{i l} \delta_{k j} \varepsilon(i \neq k) \frac{N e^{(i-k) z-\operatorname{sgn}(i-k) N z / 2}}{2 \sinh (N z / 2)}+
 }
 \\ \ \\
 \displaystyle{
+N e^{(i-j) z} \delta_{i+k, j+l}(\varepsilon(i<j<k)-\varepsilon(k<j<i))+N e^{-N \Lambda} \delta_{i+k, j+l+N}\left(e^{-j z} \delta_{i N}-e^{l z} \delta_{k N}\right)\,.
 }
  \end{array}
  \eq
The next coefficient $m_{12}(z)$ in the quasi-classical limit (\ref{a20}) is also used in what follows:
  \beq\label{a25}
  \begin{array}{c}
      \displaystyle{
m_{i j, k l}(z)=\delta_{i j} \delta_{k l} \delta_{i k} \frac{N^{2}}{12}+\delta_{i j} \delta_{k l} \varepsilon(i \neq k)\left(\frac{(i-k)^{2}}{2}+\frac{N^{2}}{12}-\frac{N}{2}|i-k|\right)+
 }
 \\ \ \\
 \displaystyle{
+N(j-k) e^{(i-j) z} \delta_{i+k, j+l}\Big(\varepsilon(i<j<k)-\varepsilon(k<j<i)\Big)-
 }
 \\ \ \\
 \displaystyle{
-N e^{-N \Lambda} \delta_{i+k, j+l+N}\left(l e^{-j z} \delta_{i N}+j e^{l z} \delta_{k N}\right)\,.
 }
  \end{array}
  \eq
Some more explicit formulae are presented in the Appendix.

$R$-matrix satisfies the quantum Yang-Baxter equation (\ref{a07}). But it is even more important for
our purpose that it satisfies the associative Yang-Baxter equation\footnote{In fact, any solution of
(\ref{a11}) with the properties (\ref{a17})-(\ref{a18}) is also a solution of (\ref{a07}).} (\ref{a11}).
This statement follows from the classification \cite{Pol2,T} of trigonometric solutions of (\ref{a11}). See
also \cite{KrZ} for a review.

\section{Calogero-Moser-Sutherland model and trigonometric top}
\setcounter{equation}{0}

Here we briefly review the relation between the finite-dimensional Calogero-Moser-Sutherland model
(\ref{a01}) and the
trigonometric top. More details can be found in \cite{KrZ}, where this relation was described at the
level of relativistic models, i.e. between the trigonometric Ruijsenaars-Schneider model and
the relativistic (the one governed by quadratic $r$-matrix structure) trigonometric top.

\paragraph{The Calogero-Moser-Sutherland model.} The Lax pair has the form:
  \beq\label{a26}
  \begin{array}{c}
      \displaystyle{
L^{\hbox{\tiny{CSM}}}_{i j}(z)=
 }
 \\ \ \\
 \displaystyle{
=\delta_{i j} \left[-p_i+\frac{1}{N}\sum\limits_{k=1}^N p_{k} + \frac{c }{2} \coth \left(\frac{N z}{2} \right)\right]
+\left(1-\delta_{i j}\right) \frac{c}{2}  \left[\coth  \left(\frac{q_j-q_i}{2}\right)+\coth \left(\frac{N z}{2} \right)\right]
 }
  \end{array}
  \eq
  and
  \beq\label{a27}
  \begin{array}{c}
      \displaystyle{
M^{\hbox{\tiny{CSM}}}_{i j}(z)=-\delta_{i j} \sum\limits_{k:k\neq i} \frac{c}{4\sinh^2(\frac{q_i-q_k}{2})}
+\left(1-\delta_{i j}\right) \frac{c}{4\sinh^2(\frac{q_i-q_j}{2})}\,.
 }
  \end{array}
  \eq
The equations of motion generated by the Hamiltonian (\ref{a01}) are represented in the  Lax form (\ref{a04}).
Hereinafter we work in the center of mass frame, i.e.
  \beq\label{a28}
  \begin{array}{c}
      \displaystyle{
\sum\limits_{k=1}^N q_k=0\,.
 }
  \end{array}
  \eq
\paragraph{Trigonometric top.} It was proved in \cite{LOZ16} that for any $R$-matrix
satisfying the associative Yang-Baxter equation (\ref{a11}) and the properties (\ref{a17})-(\ref{a23})
one can construct an integrable top-like model of type (\ref{a02})-(\ref{a03}). Namely,
define the Lax pair as follows:
%
  %
  \beq\label{a29}
  \begin{array}{c}
      \displaystyle{
L^{\hbox{\tiny{top}}}(z, S)=\tr_{2}\left(r_{12}(z) S_{2}\right),
 }
  \end{array}
  \eq
  \beq\label{a30}
  \begin{array}{c}
      \displaystyle{
M^{\hbox{\tiny{top}}}(z, S)=\tr_{2}\left(m_{12}(z) S_{2}\right),
 }
  \end{array}
  \eq
  where ${\tr}_{2}$ means the trace over the second tensor component and $S_2=1_N\otimes S$.
  That is, having $r$-matrix in the form
 \beq\label{a31}
 \begin{array}{c}
  \displaystyle{
 r_{12}(z)=\sum\limits_{i,j,k,l=1}^N r_{ij,kl}(z)E_{ij}\otimes E_{kl}\in\Mat^{\otimes 2}
  }
 \end{array}
\eq
 and using $\tr(E_{kl}S)=S_{lk}$, from (\ref{a29}) one gets
 \beq\label{a32}
 \begin{array}{c}
  \displaystyle{
 L^{\hbox{\tiny{top}}}(z)=\sum\limits_{i,j,k,l=1}^N r_{ij,kl}(z)S_{lk}E_{ij}\in\Mat\,.
  }
 \end{array}
\eq
 Then the Lax equation (\ref{a04}) with the Lax pair (\ref{a29})-(\ref{a30})
 provides the Euler-Arnold equations (\ref{a03}) with $J(S)$ defined as
 \beq\label{a33}
 \begin{array}{c}
  \displaystyle{
J(S)=\tr_{2}\left(m_{12}(0) S_{2}\right).
  }
 \end{array}
\eq
 Explicit expressions for (\ref{a29})-(\ref{a30}) and (\ref{a33}) coming from the $R$-matrix (\ref{a15})
 can be found in the Appendix.

\paragraph{Gauge equivalence.} It was mentioned in \cite{KrZ} that in order to
establish the gauge equivalence (\ref{a05}) one should fix the constant $\Lambda$
entering (\ref{a15}), (\ref{a24})-(\ref{a25}) and the formulae from the Appendix
as
 \beq\label{a900}
 \begin{array}{c}
  \displaystyle{
 \Lambda=\imath\pi\,.
  }
 \end{array}
\eq
Introduce the matrix
 \beq\label{a34}
 \begin{array}{c}
  \displaystyle{
 g(z,q)=\Xi(z, q)D^{-1}(q)\in\Mat\,,
  }
 \end{array}
\eq
where $\Xi(z,q)$ is the Vandermonde type matrix (except the last row)
 \beq\label{a35}
 \begin{array}{c}
  \displaystyle{
\Xi_{i j}(z, q)=e^{(i-1)\left(z+q_{j}\right)}+(-1)^{N} e^{-\left(z+q_{j}\right)} \delta_{i N}\,,
\qquad \sum\limits_{k=1}^Nq_k=0
  }
 \end{array}
\eq
and $D$ is the following diagonal matrix:
 \beq\label{a36}
 \begin{array}{c}
  \displaystyle{
D_{i j}(q)=\delta_{i j} \prod\limits_{k:k \neq i}^N\Big(e^{q_{i}}-e^{q_{k}}\Big)\,.
  }
 \end{array}
\eq
The matrix $g(z,q)$ is the intertwining matrix entering the IRF-Vertex relation (\ref{a06}) with
the non-standard trigonometric $R$-matrix (\ref{a15}) in the r.h.s of (\ref{a06}), see \cite{AHZ}.

Then the gauge equivalence (\ref{a05}) holds true, where on the ''top'' side we deal with the model
corresponding to the minimal coadjoint orbit. The latter means that the matrix $S$ is of rank one.

Then the relation (\ref{a05}) is valid with the following change of variables:
 \beq\label{a37}
 \begin{array}{c}
  \displaystyle{
S_{i j}(p,q,c)= \frac{(-1)^{j}\sigma_{j}(e^{q})}{N} \sum_{m=1}^{N}  \frac{{\ti p}_{m}\Big(e^{(i-1) q_{m}}+(-1)^{N} \delta_{i N} e^{-q_{m}}\Big)- c N \delta_{i N} (-1)^{N} e^{-q_{m}}}{\prod\limits^N_{l:\,l \neq m}\Big( e^{q_{m}}-e^{q_{l}}\Big)}\,,
  }
 \end{array}
\eq
  where
 \beq\label{a38}
 \begin{array}{c}
  \displaystyle{
{\ti p}_m=-p_{m}+(i-1)c-\frac{c}{2} \sum\limits_{l:\,l \neq m}^{N} \coth \Big( \frac{q_m-q_l}{2}\Big)
  }
 \end{array}
\eq
and $\sigma_{j}(e^{q})=\sigma_j(e^{q_1},...,e^{q_N})$ are elementary symmetric functions of the variables
$e^{q_i}$, that is
 \beq\label{a39}
 \begin{array}{c}
  \displaystyle{
\prod_{k=1}^{N}\left(\zeta-e^{x_{k}}\right)=\sum_{k=0}^{N}(-1)^{k} \zeta^{k} \sigma_{k}(e^{x})\,.
  }
 \end{array}
\eq
It is easy to see from (\ref{a37}) that the matrix $S$ is indeed a matrix of rank one.
The set of its eigenvalues is as follows:
 \beq\label{a40}
 \begin{array}{c}
  \displaystyle{
{\rm Spec}(S)=(0,...,0, c)\,,
  }
 \end{array}
\eq
so that $\tr(S)=c$ and
 \beq\label{a41}
 \begin{array}{c}
  \displaystyle{
S^2=  c S\,.
  }
 \end{array}
\eq
One can show that the Poisson brackets $\{S_{ij}(p,q,c),S_{kl}(p,q,c)\}$ between elements of the matrix (\ref{a37})
computed by means of the canonical brackets (\ref{a011}) reproduce the linear Poisson-Lie brackets (\ref{a021}),
that is the map between two models is Poisson (canonical).

Similar results are known for the rational and elliptic models including relativistic models, where on the many-body side one deals with the Ruijsenaars-Schneider model, and the relativistic top is described
by quadratic Poisson brackets of Sklyanin type. See \cite{LOZ,AASZ,LOZ14,KrZ,ZZ} for details.

\section{1+1 Landau-Lifshitz equations from $R$-matrices}
\setcounter{equation}{0}
In the field case the dynamical variables are again arranged into matrix
 \beq\label{a411}
 \begin{array}{c}
  \displaystyle{
 S=\sum\limits_{i,j=1}^N E_{ij}S_{ij}\,,\quad S_{ij}=S_{ij}(t,x)\,.
 }
 \end{array}
 \eq
In what follows we use short notations for partial derivatives:
 \beq\label{a561}
 \begin{array}{c}
  \displaystyle{
\p_x\varphi(t,x)=\varphi_x\,,\quad \p_t\varphi(t,x)=\varphi_t\,.
  }
 \end{array}
\eq
Having solution of the associative Yang-Baxter equation (\ref{a11}), which satisfies also the properties
(\ref{a17})-(\ref{a23}), one can derive a wide set of identities for the coefficients of expansions
(\ref{a20})-(\ref{a22}). The first example (see \cite{LOZ16}) is given by the relation
 \beq\label{a42}
 \begin{array}{c}
  \displaystyle{
 [m_{13}(z),r_{12}(z)]=[r_{12}(z),m_{23}(0)]-[\p_z m_{12}(z),P_{23}]
  +[m_{12}(z),r_{23}^{(0)}]+[m_{13}(z),r_{23}^{(0)}]\,.
 }
 \end{array}
 \eq
It leads to the construction of the Lax pair for the integrable top (\ref{a29})-(\ref{a33}).
The next identity is of the form:
 \beq\label{a43}
  \begin{array}{c}
  \displaystyle{
  r_{12}(z)r_{13}(z)=r_{23}^{(0)}r_{12}(z)-r_{13}(z)r_{23}^{(0)}
  -\p_z r_{13}(z)P_{23}+m_{12}(z)+m_{23}(0)+m_{13}(z)\,.
 }
 \end{array}
 \eq
Using (\ref{a42})-(\ref{a43}) the following construction of $U$--$V$ pair was suggested in \cite{AtZ2}.
Consider the pair of matrices:
 \beq\label{a44}
  \begin{array}{c}
  \displaystyle{
 U^{\hbox{\tiny{LL}}}(z)=L(S,z)=\tr_2\Big(r_{12}(z)S_2\Big)\in\Mat
 }
 \end{array}
 \eq
and\footnote{The last one constant term $(Nc^2/12)1_N$ in (\ref{a45}) is cancelled out
from the Zakharov-Shabat equation (\ref{a08}). It is written here for the exact matching
in the gauge equivalence (\ref{a12}).}
 \beq\label{a45}
  \begin{array}{c}
  \displaystyle{
 V^{\hbox{\tiny{LL}}}(z)=V_1(z)-c V_2(z)-\frac{Nc^2}{12}\,1_N\in\Mat\,,
 }
 \end{array}
 \eq
where
 \beq\label{a46}
  \begin{array}{c}
  \displaystyle{
 V_1(z)=-c\p_z L(S,z)+L(E(S)S,z)\,,
 }
 \\ \ \\
   \displaystyle{
 V_2(z)=L(T,z)\,.
 }
 \end{array}
 \eq
Here we use the notation
 \beq\label{a47}
  \begin{array}{c}
  \displaystyle{
E(S)=\tr_{2}\left(r_{12}^{(0)} S_{2}\right)\in\Mat\,,
 }
 \end{array}
 \eq
see (\ref{a912})-(\ref{a914}) in the Appendix. The matrix $T$ entering the definition of $V_2(z)$
is defined as
 \beq\label{a48}
  \begin{array}{c}
  \displaystyle{
T=-\frac{k}{c^2}\,[S,S_x]\,,\quad S_x=\p_x S\,.
 }
 \end{array}
 \eq
It solves equation
 \beq\label{a49}
  \begin{array}{c}
  \displaystyle{
-kS_x=[S,T]
 }
 \end{array}
 \eq
if $S$ satisfies the condition (\ref{a41}). In what follows we assume that $S$ is a rank one matrix,
so that
 \beq\label{a50}
  \begin{array}{c}
  \displaystyle{
S^2=\tr(S)S\,,\quad c=\tr(S)
 }
 \end{array}
 \eq
 in (\ref{a41}). In this case a set of additional relations appear. In particular, $SE(S)=0$ (see \cite{AtZ2}).
 Finally, the Zakharov-Shabat equation is equivalent to the following
 Landau-Lifshitz type equation:
 \beq\label{a51}
  \begin{array}{c}
  \displaystyle{
 \partial_{t} S = 2 c [S, J(S)] + \frac{k^{2}}{c} [S, S_{xx}]-2 k [S, E(S_{x})]\,.
 }
 \end{array}
 \eq
In the $N=2$ case $E(S)=0$ and the last term vanishes.

Equation (\ref{a51}) is Hamiltonian.
The Poisson brackets are defined similarly to (\ref{a021}) but for the loop algebra:
 \beq\label{a52}
 \begin{array}{c}
  \displaystyle{
 \left\{S_{i j}(x), S_{k l}(y)\right\}=\frac{1}{N}\,\Big(S_{i l} \delta_{k j}-S_{k j} \delta_{i l}\Big)
 \delta(x-y)\,.
  }
 \end{array}
\eq
The Hamiltonian
 \beq\label{a53}
 \begin{array}{c}
  \displaystyle{
  \mH^{\hbox{\tiny{LL}}}=\oint d x H^{\hbox{\tiny{LL}}}(x)
  }
 \end{array}
\eq
with the
density
 \beq\label{a54}
 \begin{array}{c}
  \displaystyle{
  H^{\hbox{\tiny{LL}}}(x)=N c \operatorname{tr}(S J(S))- \frac{N k^{2}}{2 c} \operatorname{tr}\left(\partial_{x} S \partial_{x} S\right)+ N k \operatorname{tr}\left(\partial_{x} S E(S)\right)
  }
 \end{array}
\eq
provides the equation (\ref{a51}) through the equation $\p_t S=\{S,\mH^{\hbox{\tiny{LL}}}\}$ and the brackets
(\ref{a52}).

\section{1+1 Calogero-Moser-Sutherland field theory}
\setcounter{equation}{0}

Following \cite{Krich22} let us introduce the $U$--$V$ pair with spectral parameter for 1+1 field generalization of the
Calogero-Moser-Sutherland model.
Introduce the set of functions
 \beq\label{a55}
 \begin{array}{c}
  \displaystyle{
\alpha_{i}^{2}=k q_{i x}-c\,,\quad i=1,...,N\,.
  }
 \end{array}
\eq
and a function
 \beq\label{a56}
 \begin{array}{c}
  \displaystyle{
\kappa=-\frac{1}{N c} \sum\limits_{l=1}^{N} p_{l}\left(c-k q_{l x}\right)=
\frac{1}{N c} \sum\limits_{l=1}^{N} p_{l}\al_l^2\,.
  }
 \end{array}
\eq
The $U$--$V$ pair is as follows:
 \beq\label{a57}
 \begin{array}{c}
  \displaystyle{
U^{\hbox{\tiny{2dCMS}}}_{i j}=\delta_{i j} \left[-p_i + \frac{1}{N} \sum_{k=1}^{N} p_{k} -  \frac{\alpha_{i}^{2} }{2} \coth \left(\frac{N z}{2} \right)-\frac{k \alpha _{i x}}{\alpha _i}\right]-
  }
  \\ \ \\
  \displaystyle{
-\left(1-\delta_{i j}\right) \left[\coth  \left(\frac{q_j-q_i}{2}\right)+\coth \left(\frac{N z}{2} \right)\right]\frac{\alpha_{j}^2}{2}
  }
 \end{array}
\eq
and
 \beq\label{a58}
 \begin{array}{c}
  \displaystyle{
V^{\hbox{\tiny{2dCMS}}}_{i j}= \delta_{i j}\left[-\frac{q_{i t}}{2} \coth \left( \frac{N z}{2} \right)- \frac{Nc\alpha_{i}^{2}}{4} \left(\frac{1}{\sinh^2( \frac{N z}{2})} +\frac{1}{3}\right) -\frac{\alpha_{i t}}{\alpha_{i}}+\widetilde{m}_{i}^{0} -\frac{1}{N}\sum_{l=1}^N \tilde{m}_{l}^{0}\right]-
  }
  \\ \ \\
  \displaystyle{
-\left(1-\delta_{i j}\right) \Bigg[ \frac{N c}{2} \coth \left(\frac{N z}{2} \right) \left(\coth \left(\frac{q_j-q_i}{2} \right)+  \coth \left(\frac{N z}{2} \right)\right) +
\frac{N c}{2 \sinh^2\left( \frac{q_{i}-q_{j}}{2} \right)}-
  }
  \\ \ \\
  \displaystyle{
  -\widetilde{m}_{i j} \left( \coth \left(\frac{q_j-q_i}{2} \right)+  \coth \left(\frac{N z}{2} \right)\right)\Bigg]\frac{\alpha_{j}^{2}}{2}\,,
  }
 \end{array}
\eq
where the set of functions $\tilde{m}_{i}^{0}$ in the diagonal part of $V$-matrix is of the form:
 \beq\label{a59}
 \begin{array}{c}
  \displaystyle{
\tilde{m}_{i}^{0}=p_{i}^{2}+\frac{k^2 \alpha_{i x x}}{\alpha_{i}}+2 \kappa p_{i}-\sum_{l:l \neq i}^N
\Bigg[\frac{\left(2 \alpha_{l}^{4}+\alpha_{i}^{2}\alpha_{l}^{2}\right)}{4} \left(\frac{1}{\text{sinh}^2\left( \frac{q_{i}-q_{l}}{2} \right)} +\frac{1}{3}\right)+
  }
 \end{array}
\eq
$$
+ 2 k \alpha_{l} \alpha_{l x}
\left(\coth \left( \frac{q_{i}-q_{l}}{2} \right) -\frac{(q_{i}-q_{l})}{6}\right)\Bigg]\,.
$$
The off-diagonal part of $V$-matrix contains the following expressions:
 \beq\label{a60}
 \begin{array}{c}
  \displaystyle{
\widetilde{m}_{i j}=p_{i}+p_{j}+2 \kappa+
  }
  \\ \ \\
  \displaystyle{
+\frac{k \alpha_{i x}}{\alpha_{i}}-\frac{k \alpha_{j x}}{\alpha_{j}}-\sum\limits_{n:n \neq i, j} ^N \frac{\alpha_{n}^{2}}{2} \left(\coth \left(\frac{q_{i}-q_{n}}{2} \right)+\coth \left(\frac{q_{n}-q_{j}}{2}\right)-\coth \left(\frac{q_{i}-q_{j}}{2}\right)\right)\,.
  }
 \end{array}
\eq
In the above formulae the notations (\ref{a561}) are used.
The statement is that the Zakharov-Shabat equation (\ref{a08}) provides the following set of equations of
motion:
 \beq\label{a61}
 \begin{array}{c}
  \displaystyle{
\p_t q_{i}=2 p_{i}\left(c-k q_{i x}\right)-\frac{2}{N c} \sum_{l=1}^{N} p_{l}\left(c-k q_{l x}\right)\left(c-k q_{i x}\right)\,,\quad i=1,...,N
  }
 \end{array}
\eq
and
 \beq\label{a62}
 \begin{array}{c}
  \displaystyle{
\p_t p_{i}=-2 k p_{i} p_{i x}+\frac{2 k}{N c}\left\{\sum_{l=1}^{N} p_{i} p_{l}\left(c-k q_{l x}\right)\right\}_{x}+k \left\{\frac{k^{3} q_{i x x x}}{2\left(c-k q_{i x}\right)}+\frac{k^{4} q_{i x x}^{2}}{4\left(c-k q_{i x}\right)^{2}}\right\}_{x}+
  }
  \\ \ \\
  \displaystyle{
+ \sum\limits_{l:l \neq i}^N\Bigg[k^{3} q_{l x x x} \left(\coth \left(\frac{\left(q_{i}-q_{l}\right)}{2} \right) - \frac{\left(q_{i}-q_{l}\right)}{6}  \right) -\frac{3 k^{2} q_{l x x} \left(c-k q_{l x}\right)}{2 \text{sinh}^2\left( \frac{q_{i}-q_{l}}{2} \right)} -
  }
  \\ \ \\
  \displaystyle{
 -\frac{3 k^{2} q_{l x x} \left(c-k q_{l x}\right)}{6}  - \frac{\left(c-k q_{l x}\right)^{3} \coth \left(\frac{\left(q_{i}-q_{l}\right)}{2} \right)}{2 ~ \text{sinh}^2\left( \frac{q_{i}-q_{l}}{2} \right)}
\Bigg]\,,\quad i=1,...,N\,.
  }
 \end{array}
\eq
The Hamiltonian description is given by
 \beq\label{a63}
 \begin{array}{c}
  \displaystyle{
  \mH^{\hbox{\tiny{2dCMS}}}=\oint d x H^{\hbox{\tiny{2dCMS}}}(x)
  }
 \end{array}
\eq
with the
density
 \beq\label{a64}
 \begin{array}{c}
  \displaystyle{
  H^{\hbox{\tiny{2dCMS}}}(x)=\sum_{i=1}^{N} p_{i}^{2} \left(c-k q_{i x}\right)-\frac{1}{N c}\left(\sum_{i=1}^{N} p_{i}\left(c-k q_{i x}\right)\right)^{2}-
  }
  \\ \ \\
  \displaystyle{
  -\sum_{i=1}^{N} \frac{k^{4} q_{i x x}^{2}}{4\left(c-k q_{i x}\right)}+\frac{k^{3}}{4}
  \sum_{i \neq j}^N\Big[q_{i x} q_{j x x}-q_{j x} q_{i x x}\Big] \left(\coth \left(\frac{q_{i}-q_{j}}{2}\right)-\frac{\left(q_{i}-q_{j}\right)}{6} \right)-
    }
  \\ \ \\
  \displaystyle{
  -\frac{1}{8} \sum_{i \neq j}^N\Bigg[\left(c-k q_{i x}\right)^{2}\left(c-k q_{j x}\right)+\left(c-k q_{i x}\right)\left(c-k q_{j x}\right)^{2}-
  }
    \\ \ \\
  \displaystyle{
-c k^{2} \left(q_{i x}-q_{j x}\right)^{2}\Bigg]  \Bigg(\frac{1}{\text{sinh}^2\left( \frac{q_{i}-q_{j}}{2} \right)} +\frac{1}{3} \Bigg)\,.
  }
 \end{array}
\eq
Equations of motion (\ref{a61})-(\ref{a62}) are reproduced as Hamiltonian equations
$\p_t f=\{f,\mH^{\hbox{\tiny{2dCMS}}}\}$ with the Poisson brackets (\ref{a09}).

It is important to mention that the above formulae are valid in the center of mass frame, i.e. the
condition (\ref{a28}) holds true in the field case as well. At the same time
the sum of momenta is not equal to zero.

\section{Gauge equivalence and change of variables}
\setcounter{equation}{0}

Introduce the function
 \beq\label{a65}
  \begin{array}{c}
  \displaystyle{
b(x,t)=\prod _{i<j}^N (e^{q_{i}}-e^{q_{j}})^{1/N} \prod _{l=1}^N\left(k q_{l,x}-c \right)^{1/(2N)}
 }
 \end{array}
 \eq
and the matrix
 \beq\label{a66}
  \begin{array}{c}
  \displaystyle{
G(z, q)=b(x,t) g(z,q)=b(x, t) \Xi(z, {q})D^{-1}(q)\in\Mat\,.
 }
 \end{array}
 \eq
Then the gauge equivalence (\ref{a12}) holds true relating the $U$--$V$ pairs (\ref{a57})-(\ref{a58})
and (\ref{a45})-(\ref{a46}) with the following change of variables:
 \beq\label{a67}
  \begin{array}{c}
  \displaystyle{
S_{i j}(x)= \frac{(-1)^{j}\sigma_{j}(e^{q}) }{N} \sum_{m=1}^{N}
 \frac{P_{m}\Big(e^{(i-1) q_{m}}+(-1)^{N} \delta_{i N} e^{-q_{m}}\Big)+ N \alpha_{m}^{2} (-1)^{N} \delta_{i N} e^{-q_{m}}}{\prod\limits_{l:\,l \neq m}^N\left( e^{q_{m}}-e^{q_{l}}\right)}\,,
 }
 \end{array}
 \eq
where the notations (\ref{a39}) for the elementary symmetric functions are used and
 \beq\label{a68}
  \begin{array}{c}
  \displaystyle{
P_{m}=-p_{m}-\frac{k \alpha _{m x}}{\alpha _m}-(i-1) \alpha_{m}^{2}+\frac{(N-2)}{2}k q_{m, x}+\frac{\alpha_{m}^{2}}{2} \sum_{l\,:l \neq m}^{N} \coth \left( \frac{q_m-q_l}{2}\right)
 }
 \end{array}
 \eq
for $m=1,...,N$.

Similarly to the finite-dimensional case the set of eigenvalues of the matrix $S$ has the form
 \beq\label{a69}
  \begin{array}{c}
  \displaystyle{
Spec(S)=(0,...,0, c)
 }
 \end{array}
 \eq
and
 \beq\label{a70}
  \begin{array}{c}
  \displaystyle{
\tr(S)= c,~~~ S^2=  c S\,.
 }
 \end{array}
 \eq
The Poisson brackets $\{S_{ij}(x),S_{kl}(y)\}$ being computed for the expressions (\ref{a67})
by means of the canonical brackets (\ref{a09}), provide the linear Poisson brackets (\ref{a52}).
That is the map between two models is the Poisson map.

The Hamiltonians of two models (\ref{a53}) and (\ref{a63})
 coincide with the change of variables (\ref{a67}):
 \beq\label{a71}
  \begin{array}{c}
  \displaystyle{
\mH^{\hbox{\tiny{2dCMS}}}=\mH^{\hbox{\tiny{LL}}}\,.
 }
 \end{array}
 \eq
In fact, the densities of the Hamiltonians (\ref{a54}) and (\ref{a64}) also coincide up to a constant:
 \beq\label{a72}
  \begin{array}{c}
  \displaystyle{
H^{\hbox{\tiny{2dCMS}}}(x)=H^{\hbox{\tiny{LL}}}(x)-\frac{N^{2} c^{3}}{12}\,.
 }
 \end{array}
 \eq
The above statements are verified by technically complicated but straightforward calculations, which
are similar to the finite-dimensional case discussed in \cite{KrZ} and \cite{AASZ}.
The coincidence of the Hamiltonians follows from the gauge equivalence since the Hamiltonians
are generated by traces of powers of monodromy of the connection $\nabla_x=k\p_x-U$. The canonicity
of the map between models can be verified directly. For this purpose one can represent the matrix $S$
in the form $S_{ij}=(1/N)a_ib_j$ and show that $\{b_i(x),a_j(y)\}=\delta_{ij}\delta(x-y)$ (for $i,j=1,...N-1$) follows from
the canonical brackets (\ref{a09}).
In this way one easily come to (\ref{a52}). This calculation
is also similar to the finite-dimensional case, see e.g. \cite{AASZ} in the rational case. Details of the proof will be given elsewhere.

%
%

\section{Appendix}
\def\theequation{A.\arabic{equation}}
\setcounter{equation}{0}

Here we collect some useful explicit expressions entering expansions of the non-standard $R$-matrix,
Lax equations and equations of motion of the trigonometric top and the corresponding Landau-Lifshitz model.
We emphasize that the below given formulae contain dependence on the arbitrary constant $\Lambda$
entering $R$-matrix (\ref{a15}). In order to use these formulae for relation to the Calogero-Moser-Sutherland
model one should fix $\Lambda=\imath \pi$ as in (\ref{a900}).

\paragraph{Expansions of the non-standard $R$-matrix.} We begin with the explicit expressions for $m_{12}(0)$ and $r_{12}^{(0)}$:
 \beq\label{a901}
 \begin{array}{c}
  \displaystyle{
m_{i j, k l}(0)=\delta_{i j} \delta_{k l} \delta_{i k} \frac{N^{2}}{12}+\delta_{i j} \delta_{k l} \varepsilon(i \neq k)\left(\frac{(i-k)^{2}}{2}+\frac{N^{2}}{12}-\frac{N}{2}|i-k|\right)+
  }
  \\ \ \\
   \displaystyle{
   +N(j-k) \delta_{i+k, j+l}\Big(\varepsilon(i<j<k)-\varepsilon(k<j<i)\Big)-N e^{-N \Lambda} \delta_{i+k, j+l+N}\left(l \delta_{i N}+j \delta_{k N}\right)
   }
 \end{array}
\eq
and
 \beq\label{a902}
 \begin{array}{c}
  \displaystyle{
r_{i j, k l}^{(0)}=\left(\delta_{i j} \delta_{k l} \varepsilon(i \neq k)+\delta_{i l} \delta_{k j} \varepsilon(i \neq k)\right)\left((i-k)-\frac{N \operatorname{sgn}(i-k)}{2}\right)+
  }
  \\ \ \\
   \displaystyle{
+N \delta_{i+k, j+l}\Big(\varepsilon(i<j<k)-\varepsilon(k<j<i)\Big)+N e^{-N \Lambda} \delta_{i+k, j+l+N}\left(\delta_{i N}-\delta_{k N}\right)\,.
   }
 \end{array}
\eq

\paragraph{$L$-matrix.}
The Lax matrix $L^{\hbox{\tiny{top}}}(z, S)$ following from (\ref{a29}) has the following form. Its diagonal
part:
 \beq\label{a903}
 \begin{array}{c}
  \displaystyle{
L_{ii}^{\hbox{\tiny{top}}}(z,S)=\sum _{k=1}^{i-1} \frac{(2 i-2 k-N)}{2} S_{k k} +\sum _{k=i+1}^N \frac{(2 i-2 k+N)}{2} S_{k k} +\frac{N S_{i i}}{2} \coth \left(\frac{N z}{2}\right)\,.
   }
 \end{array}
\eq
Next, for $i<j$:
 \beq\label{a904}
 \begin{array}{c}
  \displaystyle{
L_{ij}^{\hbox{\tiny{top}}}(z,S)=N
\sum\limits_{k=j+1}^N e^{z (i-j)} S_{i-j+k,k}+\frac{N S_{i j} e^{z (i-j)+\frac{N z}{2}}}{2 \sinh (N z / 2)}
   }
 \end{array}
\eq
and
for $i>j$ we have:
 \beq\label{a905}
 \begin{array}{c}
  \displaystyle{
L_{ij}^{\hbox{\tiny{top}}}(z) =N e^{-N \Lambda } \delta _{i N} \sum _{k=j+1}^N e^{-j z} S_{k-j,k}-
 }
 \\ \ \\
   \displaystyle{
   -N \sum _{k=1}^{j-1} e^{z (i-j)} S_{i-j+k,k}
-  N S_{i-j,N} e^{z (i-j)-\Lambda  N}+\frac{N S_{i j} e^{z (i-j)-\frac{N z}{2}}}{2 \sinh (N z / 2)}\,.
   }
 \end{array}
\eq

\paragraph{$M$-matrix.} The $M$-matrix defined through (\ref{a30}) and (\ref{a25})
 has the following explicit form.
The diagonal part:
 \beq\label{a906}
 \begin{array}{c}
  \displaystyle{
M_{ii}^{\hbox{\tiny{top}}}(z,S)=\sum _{k=1}^{i-1} \frac{1}{12} S_{k k} \left(6 i^2-12 i k-6 i N+6 k^2+6 k N + N^2\right) +
 }
 \\ \ \\
   \displaystyle{
 + \sum _{k=i+1}^N \frac{1}{12} S_{k k} \left(6 i^2-12 i k+6 i N+6 k^2-6 k N+N^2\right)+\frac{1}{12} N^2 S_{i i}\,.
   }
 \end{array}
\eq
For $i<j$:
 \beq\label{a907}
 \begin{array}{c}
  \displaystyle{
M_{ij}^{\hbox{\tiny{top}}}(z,S)=N \sum _{k=j+1}^N (j-k) e^{z (i-j)} S_{i-j+k,k}
   }
 \end{array}
\eq
and
for $i>j$:
 \beq\label{a908}
 \begin{array}{c}
  \displaystyle{
M_{ij}^{\hbox{\tiny{top}}}(z,S) =N e^{-N \Lambda } \delta _{i N} \sum _{k=j+1}^N (j - k) e^{-j z} S_{k-j,k}
-N \sum _{k=1}^{j-1} (j - k) e^{z (i-j)} S_{i-j+k,k}-
 }
 \\ \ \\
   \displaystyle{
-  N j S_{i-j,N} e^{z (i-j)-\Lambda  N}\,.
   }
 \end{array}
\eq

\paragraph{$J(S)$-matrix.} The matrix $J(S)$ is defined through (\ref{a33}) and (\ref{a901}).
Its diagonal part is as follows:

 \beq\label{a909}
 \begin{array}{c}
  \displaystyle{
J_{i i}(S)=\sum _{k=1}^{i-1} \frac{1}{12} S_{k k} \left(6 i^2-12 i k-6 i N+6 k^2+6 k N + N^2\right) +
 }
 \\ \ \\
   \displaystyle{
 + \sum _{k=i+1}^N \frac{1}{12} S_{k k} \left(6 i^2-12 i k+6 i N+6 k^2-6 k N+N^2\right)+\frac{1}{12} N^2 S_{i i}\,.
   }
 \end{array}
\eq
For $i<j$ :
 \beq\label{a910}
 \begin{array}{c}
  \displaystyle{
J_{i j}(S)=N \sum _{k=j+1}^N (j-k) S_{i-j+k,k}
   }
 \end{array}
\eq
and
for $i>j$ :
 \beq\label{a911}
 \begin{array}{c}
  \displaystyle{
J_{i j}(S) =N e^{-N \Lambda  } \delta _{i N} \sum _{k=j+1}^N (j-k) S_{k-j,k}-N \sum _{k=1}^{j-1} (j-k) S_{i-j+k,k}-j N e^{- N \Lambda } S_{i-j,N}\,.
   }
 \end{array}
\eq

\paragraph{$E(S)$-matrix.} The matrix $E(S)$ is defined as in (\ref{a47}) with $r_{12}^{(0)}$ given
in (\ref{a902}). Its diagonal part has the form:
 \beq\label{a912}
 \begin{array}{c}
  \displaystyle{
E_{i i}(S)=\sum _{k=1}^{i-1} \frac{1}{2} S_{kk} (2 i-2 k-N)+\sum _{k=i+1}^N \frac{1}{2} S_{kk} (2 i - 2 k+N)\,.
   }
 \end{array}
\eq
The off-diagonal part with
$i<j$ we have:
 \beq\label{a913}
 \begin{array}{c}
  \displaystyle{
E_{i j}(S)=N \sum _{k=j+1}^N S_{i-j+k,k}+\left(i-j+\frac{N}{2}\right) S_{i j}
   }
 \end{array}
\eq
and finally for
 $i>j$:
 \beq\label{a914}
 \begin{array}{c}
  \displaystyle{
E_{i j}(S) = N e^{-N \Lambda } \delta _{i N} \sum _{k=j+1}^N S_{k-j,k}-N \sum _{k=1}^{j-1} S_{i-j+k,k}-N e^{- N \Lambda } S_{i-j, N}+\left(i-j-\frac{N}{2}\right) S_{i j}\,.
   }
 \end{array}
\eq
 %






\begin{small}

\end{small}

\end{document}